# Adaptive modulation in the Ni$_2$Mn$_{1.4}$In$_{0.6}$ magnetic shape memory Heusler alloy

P. Devi[1], Sanjay Singh[1,2*], B. Dutta[3], K. Manna[1], S. W. D'Souza[1], Y. Ikeda[3] E. Suard[4], V. Petricek[5], P. Simon[1], P. Werner[6], S. Chadhov[1], Stuart S. P. Parkin[6], C. Felser[1], and D. Pandey[2]

[1] Max Planck Institute for Chemical Physics of Solids, Nöthnitzer Strasse 40, D-01187 Dresden, Germany

[2] School of Materials Science and Technology, Indian Institute of Technology (Banaras Hindu University), Varanasi-221005, India.

[3] Max-Planck-Institut für Eisenforschung Max-Planck-Strasse 1, 40237, Düsseldorf, Germany

[4] Institut Laue-Langevin, BP 156, 38042 Grenoble Cedex 9, France

[5] Institute of Physics ASCR, Department of Structure Analysis, Na Slovance 2, 182 21 Praha, Czech Republic

[6] Max Planck Institute of Microstructure Physics, 06120 Halle, Germany

**Abstract**

The origin of incommensurate structural modulation in Ni-Mn based Heusler type magnetic shape memory alloys (MSMAs) is still an unresolved issue inspite of intense focus on this due to its role in the magnetic field induced ultra-high strains. In the archetypal MSMA Ni$_2$MnGa, the observation of 'non-uniform displacement' of atoms from their mean positions in the modulated martensite phase, premartensite phase and charge density wave as well as the presence of phason broadening of satellite peaks have been taken in support of the electronic instability model linked with a soft acoustic phonon. We present here results of a combined high resolution synchrotron x-ray powder diffraction (SXRPD) and neutron powder diffraction (NPD) study on Ni$_2$Mn$_{1.4}$In$_{0.6}$ using (3+1)D superspace group approach, which reveal not only uniform atomic displacements in the modulated structure of the martensite phase with physically acceptable ordered magnetic moments in the antiferromagnetic phase at low temperatures but also the absence of any premartensite phase and phason broadening of the satellite peaks. Our HRTEM studies and first principles calculations of the ground state also support uniform atomic displacements predicted by powder diffraction studies. All these observations suggest that the structural modulation in the martensite phase of $Ni_2Mn_{1.4}In_{0.6}$ MSMA can be explained in terms of the adaptive phase model. The present study underlines the importance of superspace group analysis using complimentary SXRPD and NPD in understanding the physics of the origin of modulation as well as the magnetic and the modulated ground states of the Heusler type MSMAs. Our work also highlights the fact that the mechanism responsible for the origin of modulated structure in different Ni-Mn based MSMAs may not be universal and it must be investigated thoroughly in different alloy compositions.

Magnetic shape memory Heusler alloys (MSMAs) in the Ni-Mn-X (X= Ga, In, Sn) system have enormous potential for technological applications due to a rich variety of properties ranging from generation of extremely large magnetic field induced strain (~10%) to pronounced magnetocaloric and barocaloric effects, large magnetoresistance, anomalous Hall effect and large exchange bias[1-6]. The technologically significant physical properties of these alloys are intimately linked with the coupling between structural and magnetic degrees of freedom below the magneto-structural (martensite) phase transition. Since the modulated crystal structure of martensite phase due to its low detwinning stress is known to play an important role in deciding the response of these alloys to the external magnetic field, there is currently a lot of interest in understanding the origin of the modulated structure of the martensite phase itself [7-17]. Two different models have been proposed in the literature for the origin of modulation in the MSMAs. The first one is the adaptive phase model in which the modulated structure is considered as a nanotwinned state of the Bain distorted phase, which maintains the invariance of the habit plane between the high temperature austenite and the low temperature martensite phases[14, 15]. Commensurate modulated structure, uniform atomic displacements and absence of any premartensite phase transition as well as phason strains are the key manifestations of this model. The second model, based on charge density wave (CDW) coupled to a soft transverse acoustic (TA2) mode [18-20], can, on the other hand, explain non-uniform atomic displacements, incommensurate nature of modulation and existence of premartensite phase as well as phasons [10-12, 21, 22]. It has been proposed that the formation of discommensurations in the form of stacking faults and antiphase boundaries can in principle result into an average incommensurate modulated structure even for the adaptive phase model [14-16]. However, one of the key features for distinguishing between the two models for the origin of modulation is the identification of the nature (uniform versus non-uniform) of the atomic displacements.

Recently we have shown that the origin of modulation in $Ni_2MnGa$ shape memory Heusler alloy cannot be explained within the framework of adaptive phase model as the modulated structure has non-uniform atomic displacements [12]. Also the fact that the incommensurate martensite phase results from an incommensurate premartensite phase and not directly from the austenite phase does not support the adaptive phase model [11]. The presence of phasons [23] and the broadening of the superlattice peaks due to phason strains [12] in $Ni_2MnGa$ also goes against the concept of adaptivity. On the other hand, the Ni-Mn-In MSMAs do not exhibit the premartensite (precursor) phase formation as the austenite phase transforms directly to the martensite phase. This suggests that these alloys may be model systems for investigating the applicability of the adaptive phase model for structural modulation through a careful analysis of the structure of the martensite phase. Here, we present results of Rietveld analysis of high resolution synchrotron x-ray powder diffraction (SXRPD) and neutron powder diffraction (NPD) data on a Ni-Mn-In alloy, $Ni_2Mn_{1.4}In_{0.6}$, using (3+1) D superspace group approach. Our analysis reveals that the modulated structure of the martensite phase of $Ni_2Mn_{1.4}In_{0.6}$ involves uniform displacements of atoms with respect to their positions in the Bain distorted basic cell. This conclusion is well supported by high-resolution transmission electron microscopic studies as well as first principles calculations of the ground state of these alloys.. All these observations comprehensively rule out the applicability of the CDW based soft mode model and support the adaptive modulation model for $Ni_2Mn_{1.4}In_{0.6}$. Further, in contrast to $Ni_2MnGa$, Rietveld analysis of the neutron powder diffraction pattern reveals antiferromagnetic (AFM) correlations for Mn spins at two different crystallographic positions of the martensite phase with full and partial Mn occupancies. The AFM ordering is further supported by isothermal magnetization measurements that reveal a double hysteresis loop due to a spin flop transition induced by a very low magnetic field of 0.05 T at 2K.

The details of sample preparation, measurements (magnetization, SXRPD, NPD and high resolution transmission electron microscopy (HRTEM)), Rietveld refinements and first principles calculations are given in the Supplemental Material[24]). The low field (500 Oe) magnetization curves of $Ni_2Mn_{1.4}In_{0.6}$ recorded under ZFC, FCC and FCW conditions at 0.05 T in the temperature range 2-400 K are shown in Fig.1a. It reveals a sharp jump in magnetization at the Curie temperature $T_c \sim 315$ K due to a ferromagnetic (FM) transition, followed by a decrease in magnetization at the first order martensite transition temperature $T_M \sim 295$ K. The bifurcation of the ZFC and FCC curves below T~145 K is in agreement with earlier reports [3] where it has been attributed to the coexistence of antiferromagnetic and ferromagnetic exchange interactions in the martensite phase [37]. Our isothermal magnetization M(H) plot at 2 K (see Fig. 1b) indicates typical antiferromagnetic ground state with a spin flop transition occurring at a very low magnetic field of ±0.05 T leading to opening up of double hysteresis loop above this field. The fact that the spin flop transition occurs at such a low field suggests that both the FM and AFM states are nearly degenerate in the martensite phase, even though the ground state is dominated by AFM interactions.

We now turn towards the structure of the austenite and martensite phases using SXRPD patterns recorded at 350 K (austenite phase) and 235 K (martensite phase), respectively. In the first step of the structure analysis, we performed indexing of the powder diffraction patterns by LeBail technique, which refines the unit cell parameters and profile broadening functions to obtain the best fit between the observed and calculated profiles in the least squares sense for a given space group. At 350 K, all the observed Bragg peaks could be indexed well with the cubic austenite structure (space group Fm-3m) and the refined lattice parameter is found to be 6.00483(4) Å. The presence of the superstructure peaks like (111) and (200) in the SXRPD pattern (inset of Fig.2a) confirms that the structure corresponds to the ordered $L2_1$ type [38]. At 235 K, many more reflections appear and the cubic austenite peaks split into two or more peaks clearly indicating a non-cubic structure. A careful analysis of all the observed low intensity peaks revealed that the martensite structure at 235 K cannot be explained in terms of a simple Bain distorted unit cell and requires consideration of modulation of the Bain distorted unit cell as reported in other MSMAs [7, 8, 11, 39, 40]. Superspace (3+1) D formalism [41-44] is a powerful tool to investigate such complex modulated structures and we employed this formalism to investigate the structure of the modulated martensite phase in $Ni_2Mn_{1.4}In_{0.6}$. Following the superspace group formalism, the SXRPD pattern was divided into two sets of reflections: (1) main reflections corresponding to the Bain distorted basic structure and (2) satellite reflections due to the modulation whose intensity is in general much less than the intensity of the main reflections. All the main reflections corresponding to the basic structure could be indexed with a monoclinic cell with space group *I*2/m and Le-Bail refinement gave us lattice parameters as a=4.3983(1) Å, b= 5.6453(2) Å, c= 4.3379(1) Å and β= 92.572(2)$^0$. After obtaining the cell parameters for the basic structure, the full SXRPD pattern including both the main and the satellite reflections was considered for Le-Bail refinement using the superspace group formalism. The satellite reflections were indexed using a modulation wave vector q= (0, 0, 1/3) and superspace group *I*2/m(α0γ)00. Although this commensurate wave vector could index many of the satellite reflections some of the calculated satellite reflections were found to be shifted away from the observed reflection positions, as can be clearly seen in the inset of Fig.2b. Therefore the wave vector **q** was allowed to be refined and an excellent match between the observed and calculated profiles was obtained for an incommensurate modulation wave vector **q**= 0.35987(8) c*= (1/3+δ) c* (where δ=0.02653 is the degree of incommensuration), including those which could not be accounted for using the commensurate wave vector q=1/3 (see Fig.2c). This indicates that the martensite phase of $Ni_2Mn_{1.4}In_{0.6}$ has an incommensurate modulation which is 3M like (see Ref.[10] for definition of this notation). The SXRPD pattern shows 2$^{nd}$ order satellites (indicated by blue arrows in Fig.2c) which is consistent with 3M like modulation. A similar 3M (sometimes also labelled as 6M (for definitions, see Ref.[10]) modulated martensite structure has been reported for another Ni-Mn-In shape memory alloy composition with martensite transition temperature higher than the present alloy composition [17]. It is interesting to note that the peak broadening of both the main and satellite reflections could be successfully modeled using anisotropic strains as per Stephen's model without invoking 4$^{th}$ rank strain tensor for phason broadening [45, 46]. This *is in marked contrast to the situation in $Ni_2MnGa$ where phason strains had to be invoked to model the* *broadening of the satellite peaks.* [12].

So far we discussed the results of Le Bail refinements only, where the atomic positions were not refined. Now we proceed to discuss the results of Rietveld refinement, where the atomic positions and atomic modulation functions were also refined. In the Rietveld refinement, Ni, Mn and In atoms were considered to occupy the 4$h$ (0.5 0.25 0), 2$a$ (0 0 0) and 2$d$ (0 0.5 0) Wyckoff positions, respectively, of the basic structure. The excess Mn atoms occupy the In site (2$d$). Further, the deviation of atoms $u(\bar{x}4)$ from the average structure due to the modulation was modelled using a harmonic atomic modulation function:

$$uj(\bar{x}4) = \sum_{n=1}^{\infty}[A_n^j \sin(2\pi n\bar{x}4) + B_n^j \cos(2\pi n\bar{x}4)] \quad (1),$$

where $A_n^j$ and $B_n^j$ are the Fourier amplitudes of the displacement modulation of the j$^{th}$ atom while "n" is the order of the Fourier series, which is taken as equivalent to the highest order of the satellite reflections observed [47] which is n=2 in the present case. In the Rietveld refinement, the amplitudes of the atomic modulation function were refined without any constraints for different atomic sites as per the non-uniform displacement model used in the refinements of Ni$_2$MnGa [12] system. While this refinement yields reasonable fit between the observed and the calculated peak profiles (see Fig.3a), the calculated interatomic distances are physically unrealistic for the Ni-Mn-In family of intermetallic compounds/alloys. For example, the sum of the atomic radii (1.25 Å for Ni, 1.37 Å for Mn and 1.67 Å for In) of various pairs of atoms is always ≥ 2.5 Å, whereas the interatomic distances obtained after refinement for the 'non-uniform displacement' model are less than 2.3Å for some *t* values (see Fig.4a). One of the reasons for the physically unrealistic interatomic distances could be the presence of anti-site disorder commonly observed in Ni-Mn based Heusler alloys. This can in turn affect the amplitude of atomic modulation function, if it is not explicitly accounted for in the refinement. Since Ni and Mn have similar x-ray atomic scattering factors, anti-site disorder involving these atoms is not distinguishable by XRD data analysis. On the other hand, the scattering lengths for Ni and Mn for neutrons have opposite signs and hence neutron scattering is ideally suited for capturing Ni-Mn anti-site disorder. We therefore performed neutron powder diffraction measurements also. The Rietveld refinement for the austenite phase at room temperature confirmed the absence of any discernible anti-site disorder (for details, see Supplemental Material[24], Sec. A. III). In the next step, we therefore considered a 'uniform displacement' model for Rietveld refinement in which the amplitude of modulation for all the atomic sites were constrained to be identical. The results of Rietveld refinement for the 'uniform displacement' model is shown in Fig.3b and the corresponding atomic positions are listed in Table I. The interatomic distances obtained for the refined structure using the uniform atomic displacement model are found to be physically realistic (see Fig. 4b) and acceptable for the shape memory Heusler compounds/alloys. *As an additional check, we also carried out refinements for the 'non-uniform displacement' model using constraints on the interatomic distances so that they correspond to physically plausible values. However, these refinements converged to the values obtained for the 'uniform displacement' model.* Thus, our results suggest that the modulation in the martensite phase of $Ni_2Mn_{1.4}In_{0.6}$ involves uniform displacement of atoms, which is consistent with the adaptive phase model.

After getting the correct atomic modulation model from the analysis of SXRPD, we now proceed to discuss the magnetic structure of the martensite phase using neutron diffraction data collected at 3 K. Rietveld refinements using neutron diffraction data also support the uniform atomic displacement model (please see Supplemental Material[24], Sec. A. IV for more details). There are four possible magnetic subgroups of the nuclear superspace group $I2/m(\alpha 0\gamma)00$ (i.e., magnetic superspacegroups or mSSG) due to time reversal symmetry breaking: (i) $I2/m(α0γ)00$ (ii) $I2'/m(α0γ)00$ (iii) $I2/m'(α0γ)00$ and (iv) $I2'/m'(α0γ)00$. Of these, only (i) and (iv) allow non-zero magnetic moments. The (i) and (iv) mSSG restrict magnetic moments along the *b*-axis of the monoclinic cell. Out of these two, our Rietveld refinement (see Supplemental Material[24], Sec. A. IV) reveals that the magnetic structure can be described by $I2/m(α0γ)00$ mSSG in which the magnitude of the Mn magnetic moments of the fully occupied site (*2a*) and partially occupied site (*2d*) are equal (1.18 $μ_B$) but antiferromagnetically correlated. This is consistent with the double hysteresis loop type M(H) plot shown in Fig 1b. Thus both the magnetization and neutron results confirm that the low temperature martensite phase of $Ni_2Mn_{1.4}In_{0.6}$ contains antiferromagnetic correlations.

*Additional support for the 'uniform displacement' model was obtained through high-resolution transmission electron microscopy studies (HRTEM). The martensite phase of $Ni_2Mn_{1.4}In_{0.6}$ was observed by in-situ cooling of the sample down to 100 K. Fig.5a shows a noise filtered HRTEM image recorded along the [210] zone. In this crystal lattice projection, atoms appear as bright spots. The (001) atomic planes have an interplanner spacing of 2.1 Å. The occurrence of bright and dark horizontal bands is related to a different stacking of the (001) planes, which generate the unit cell of 3M stricture. In the martensite phase this is a stacking of six atomic planes (c = 6x 2.15 Å = 12.98 Å), which is shown in Fig. 5b. The corresponding atomic positions are listed in Table S6 of Supplemental Material[24] for the rational approximant structure of the martensite phase and correlate to the 'uniform displacement' model. In this projection, the twinning of the (001) planes is indicated by dark zig-zag line within the 3M unit cell. The experimental HRTEM image in (a) includes a region (white rectangle) of such a stacking sequence. The atom positions are consistent with the simulated positions obtained using Rietveld refined coordinates for the*

*'uniform displacement' model. It has to be mentioned that the in-situ cooling of the thin TEM samples does not always generate a perfect (4 -2) twinned structure of six atomic planes obtained for the average structure using bulk sample. The HRTEM images often show stacking faults, which locally could lead to different periodicities, say of 7 atomic planes.*

The conclusions based on the structure refinements and HRTEM studies were verified using first-principles calculations (the details of which are given in the Supplemental Material[24], Sec. B). Even though we obtained the convergence of the self-consistent calculation for the uniform as well as non-uniform atomic displacement models, many quantities (such as the Fermi energy, local magnetic moments, etc.) appeared to be unrealistic for the non-uniform displacement model. In particular, the total energy appears to be incomparably high (several hundred Ry per unit cell) for the non-uniform atomic displacement model with respect to that of the uniform atomic displacement model. Therefore, the non-uniform displacement modulation model does not appear to be realistic.

It is evident from the foregoing results of SXRPD, NPD and HRTEM investigations as well as the ab-initio calculations of the ground state that the modulated structure of $Ni_2Mn_{1.4}In_{0.6}$ Heusler alloy involves 'uniform displacement of atoms. *It is interesting to note that the experimentally observed phonon dispersion curves for $Ni_2MnGa$ reveal a dip in one of the acoustic branches($TA_2$) around wave vector q ~ (1/3 1/3 0) whereas the same acoustic branch of $Ni_2Mn_{1.4}In_{0.6}$ (~$Ni_{49.3}Mn_{34.2}In_{16.5}$) does not reveals any dip although the entire branch has rather low energy [19, 48-51]. Surprisingly, the previous first principle calculations for stoichiometric $Ni_2MnIn$ predicted similar qualitative features as that $Ni_2MnGa$ for the $TA_2$ phonon branch whereas experimentally $Ni_2MnIn$ does not undergo a martensite transition [19, 20, 48-51]. This shows* *that the first principle calculations are unable to capture the essential experimentally observed features of the phonon dispersion curves. However, the difference in the nature of the $TA_2$ acoustic branch $Ni_2Mn_{1.4}In_{0.6}$ and stoichiometric/off-stoichiometric Ni-Mn-Ga alloys in the experimental phonon dispersion curves clearly suggest that the formation of the martensite phase in In based alloy may not be mediated by phonon softening. This provides additional support to the possibility of adaptive modulation in $Ni_2Mn_{1.4}In_{0.6}$ MSMA.*

To summarise, we have critically evaluated the applicability of two existing models (electronic instability and the nano-twinning based adaptivity models) for the origin of modulation in $Ni_2Mn_{1.4}In_{0.6}$ magnetic shape memory alloy. We carried out Rietveld analysis of high resolution SXRPD and powder neutron diffraction patterns of the austenite and modulated martensite phases of $Ni_2Mn_{1.4}In_{0.6}$ using (3+1) D superspace formalism. We have considered both non-uniform and uniform displacement models of incommensurate modulation in the Rietveld refinements and shown that the nature of modulation in $Ni_2Mn_{1.4}In_{0.6}$ involves uniform atomic displacement of atoms as expected for the model based on adaptivity. This is also supported by HRTEM and ab-initio calculations. However, investigation of electronic structure using single crystal will be further useful for unambiguously proof of adaptivity in this system. Further we have shown that the magnetic structure of the martensite phase at 3K is site disordered antiferrimagnetic where Mn atoms at two different crystallographic positions are coupled antiferromagnetically. The present study underlines the importance of superspace group analysis of the diffraction data to understand the physics of modulation in magnetic shape memory Heusler alloys.

*sanjay.singh@cpfs.mpg.de

**Acknowledgments**

We thank Branton Campbell for useful discussion. S. S. thanks Alexander von Humboldt foundation, Germany for Research Fellowship and Science and Engineering Research Board of India for the award of Ramanujan Fellowship. BD acknowledges Deutsche Forschungsgemeinschaft (Germany) for the financial support within the priority programme SPP1599. DP thanks the Science and Engineering Research Board of India for the award of J.C. Bose National Fellowship. VP acknowledge support of project No. LO1603 under the Ministry of Education, Youth and Sports National sustainability program I of Czech Republic.

**Figures:**

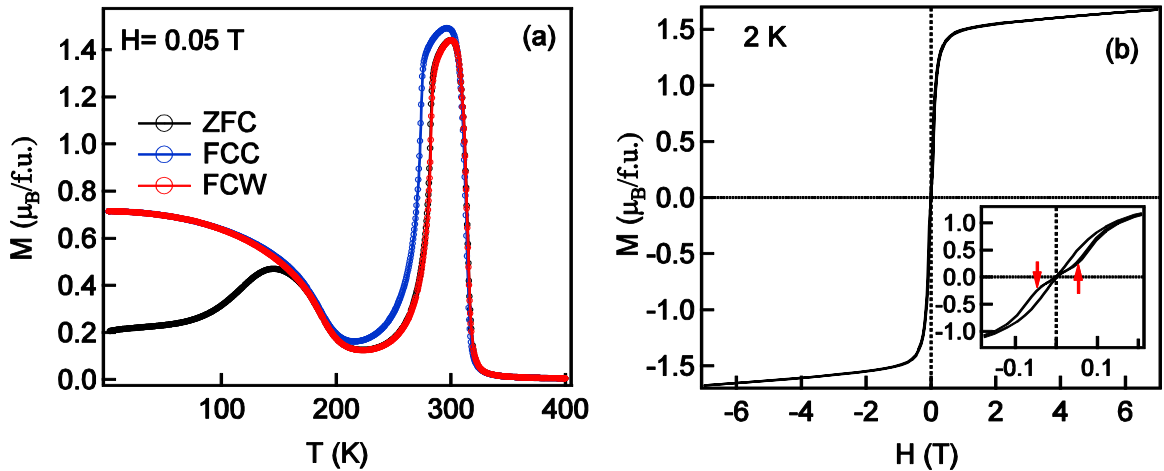

**Fig.1**:(color online) (a) Magnetization (ZFC, FC and FW) as a function of temperature at 0.05 T and (b) Magnetization as a function of field (M(H)) at 2K. Inset shows the M(H) in expanded scale, where spin flop transitions are marked by arrows.

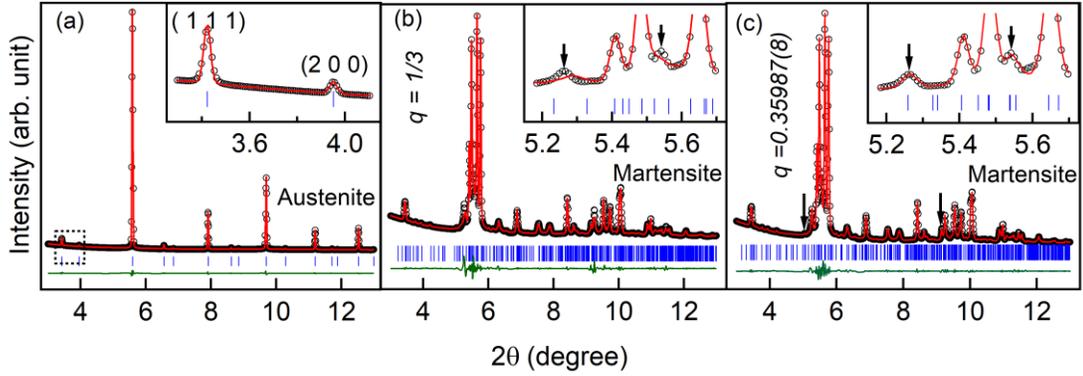

**Fig.2**: (Color online) LeBail fits for the SXRPD patterns of $Ni_2Mn_{1.4}In_{0.6}$ at (a) cubic austenite phase (350 K). Inset shows superlattice reflections related to $L2_1$ ordering (b) Martensite phase (235 K) with commensurate structure model and (c) Martensite phase (235 K) with incommensurate structure model. The insets show the fit for the main peak region ($2\theta= 5-6^0$) on an expanded scale. Arrows in (b) and (c) represent satellite reflections. The experimental data, fitted curve, and the residue are shown by circles (black), continuous line (red), and bottom-most plot (green), respectively. The tick marks (blue) represent the Bragg peak positions.

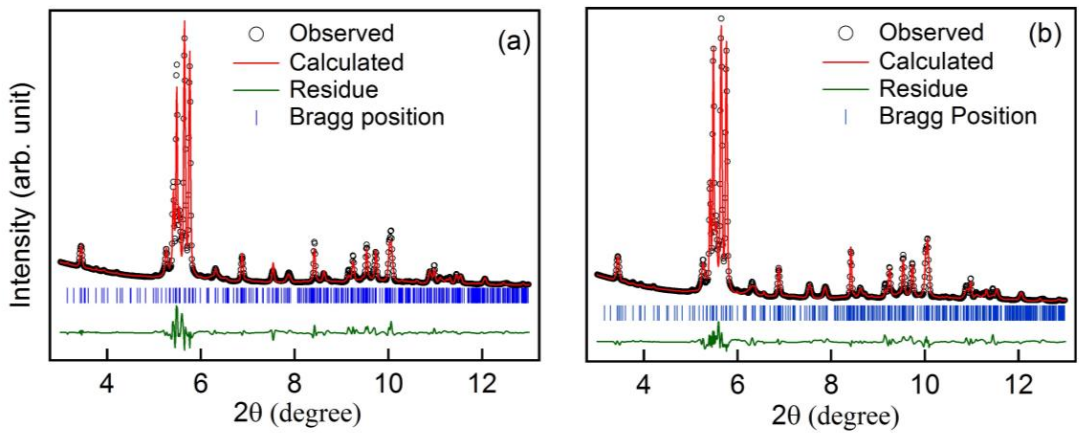

**Fig.3:** (color online) Rietveld fits for the SXRPD patterns of $Ni_2Mn_{1.4}In_{0.6}$ in martensite phase (235 K) considering (a) A non-uniform atomic displacement (electronic instability model) structure model and (b) An uniform atomic displacement (Adaptive phase) structure model. The experimental data, fitted curve, and the residue are shown by circles (black), continuous line

(red), and bottom-most plot (green), respectively. The tick marks (blue) represent the Bragg peak positions.

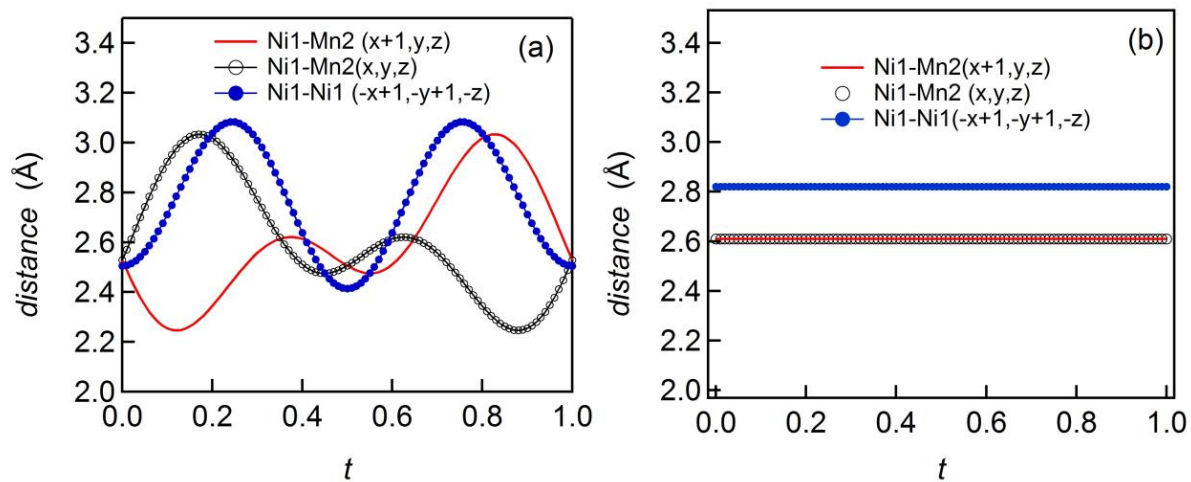

**Fig.4**: (color online) Distance (selected) as a function of *t* parameters derived from (a) Non-uniform atomic displacement model (electronic instability model) showing unphysical values (less than 2.5Å) and (b) Uniform atomic displacement model (adaptive phase model) showing values that are expected for these kind of intermetallic compounds/alloys.

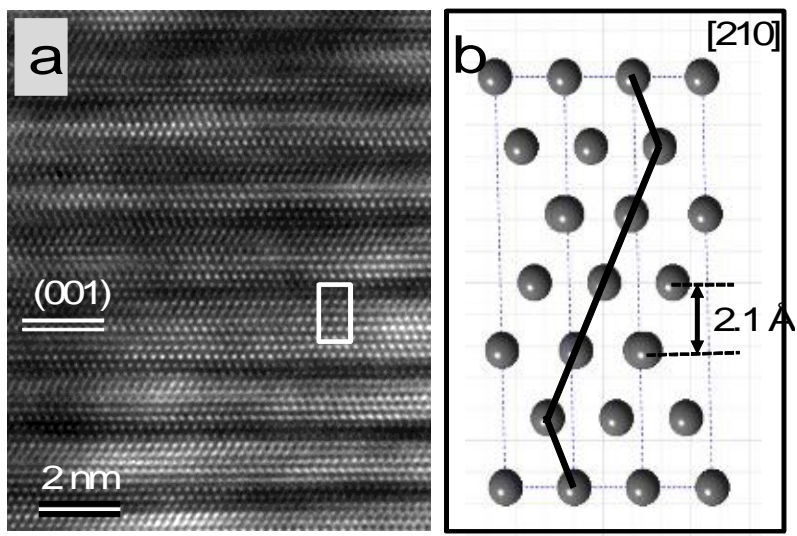

**Fig. 5.** (a) High-resolution TEM image of the $Ni_2Mn_{1.4}In_{0.6}$ crystal recorded along [210] zone axis. It represents the martensite phase (100 K). The bright spots correspond to the projected atom rows. The crystal lattice is characterized by a specific stacking of (001) basic planes (distance 2.1 Å), which can be regarded as horizontal twinned lamellae. (b) Crystal lattice model of the unit cell consisting of 6 (001) planes. The specific stacking, marked by dark lines, correspond to the uniform displacement model. Since all three kinds of atoms lie in the projection an "average color" is used. The size of such a lattice structure is marked as a rectangle in (a).

**Table I:** Atomic positions $(x, y, z)$, atomic displacement parameter $(U_{iso})$ and amplitudes $(A_1, B_1, A_2, B_2)$ of the modulation function of the modulated martensite phase (235 K) of $Ni_2Mn_{1.4}In0.6$ obtained from the Rietveld refinement of SXRPD data considering adaptive phase model.

| Atom | Wyckoff position | Modulation amplitude | x | y | z | $U_{iso}(Å^2)$ |
|---|---|---|---|---|---|---|
| Ni1 | 4h | | 0.5 | 0.25 | 0 | 0.0084(5) |
| Mn1 | 2a | | 0 | 0 | 0 | 0.0084(5) |
| In1 | 2d | | 0 | 0.5 | 0 | 0.0084(5) |
| Mn2 | 2d | | 0 | 0.5 | 0 | 0.0084(5) |
| | | $A_1$ | 0.1015(7) | 0 | 0.0018(8) | |
| | | $B_1$ | 0 | 0 | 0 | |
| | | $A_2$ | 0.0259(14) | 0 | 0.0034(19) | |
| | | $B_2$ | 0 | 0 | 0 | |

# Supplemental Material

## Adaptive modulation in the $Ni_2Mn_{1.4}In_{0.6}$ magnetic shape memory Heusler alloy

P. Devi[1], Sanjay Singh[1,2*], B. Dutta[3], K. Manna[1], S. W. D'Souza[1], Y. Ikeda[3], E. Suard[4], V. Petricek[5], P. Simon[1], P. Werner[6], S. Chadhov[1], Stuart S. P. Parkin[6], C. Felser[1], and D. Pandey[2]

[1] Max Planck Institute for Chemical Physics of Solids, NöthnitzerStrasse 40, D-01187 Dresden, Germany

[2] School of Materials Science and Technology, Indian Institute of Technology (Banaras Hindu University), Varanasi-221005, India.

[3] Max-Planck-Institut für Eisenforschung Max-Planck-Strasse 1, 40237, Düsseldorf, Germany

[4] Institut Laue-Langevin, BP 156, 38042 Grenoble Cedex 9, France

[5] Institute of Physics ASCR, Department of Structure Analysis, Na Slovance 2, 182 21 Praha, Czech Republic

[6] Max Planck Institute of Microstructure Physics, 06120 Halle, Germany

## A. Experiment and results:

Polycrystalline ingot of $Ni_2Mn_{1.4}In_{0.6}$ was prepared by melting appropriate quantities of the constituent metals of 99.99% purity under argon atmosphere using an arc furnace. The ingot was annealed in vacuum at 973 K for three days in sealed quartz ampules and then quenched into ice water. The temperature ($M(T)$) and field dependence ($M(H)$) of the magnetization curves were measured using a superconducting quantum interference device (SQUID) magnetometer. The M(T) measurements were carried out during warming after zero field cooling (ZFC), field-cooled- cooling (FC) and field-cooled- warming (FCW) conditions. Pieces cut from the ingots were mechanically ground into powder using an agate mortar and pestle and were further annealed at 773 K under high vacuum in order to remove the stresses introduced during grinding[1]. The high-resolution synchrotron powder XRD (SXRPD) measurements were performed on such powder samples at a wavelength of 0.20715 Å at P02 beamline in Petra III, Hamburg, Germany. Powder neutron diffraction (PND) pattern was also recorded using neutrons of wavelength 1.59 Å at D2B neutron diffractometer in Institut Laue-Langevin (ILL), Grenoble. The specimen was placed in a cylindrical vanadium cylinder inside a furnace for recording PND patterns at 300 K (austenite phase) and 3K (martensite phase). The powder diffraction patterns were analyzed by Le Bail and Rietveld technique using Jana2006 software package[2]. High resolution transmission electron microscopy (HRTEM) investigations were done with a TITAN transmission electron microscope and an acceleration voltage of 300 kV. To obtain the low temperature martensitic phase, a N2-cooled sample holder was used. High–resolution TEM data were analyzed by using the Digital Micrograph (DM) software (Gatancompany, USA). The high–resolution TEM images were filtered by the ASBF filter method using a script in DM created by Dave Mitchell based on the work of R. Kilaas [3] where the noisy background is subtracted. Visualization of the models was performed with the VESTA 3 software [4] Electron diffraction patterns were simulated with the program JEMS (version:3.5930U2010) [5].

**(I) Structural parameters obtained from the Rietveld refinement of SXRPD data considering non-uniform displacement (electronic instability model):**

The refined structural parameters (atomic positions and amplitude of modulation) obtained from the Rietveld refinements of SXRPD data using non-uniform atomic displacement model are shown in Table **S1**. As in this model the amplitudes of the atomic modulation function were refined without any constraints for different atomic sites therefore the amplitudes ($A_1$, $B_1$, $A_2$, $B_2$) are shown separately for each atom.

**Table S1:** Atomic positions $(x, y, z)$, amplitudes ($A_1$, $B_1$, $A_2$, $B_2$) of the modulation function, and atomic displacement parameter ($U_{iso}$) of the modulated martensite phase of $Ni_2Mn_{1.4}In0.6$ obtained from the Rietveld refinement using SXRPD data considering non-uniform displacement model.

| Atom | Wyckoff position | Modulation amplitude | x | y | z | $U_{iso}(Å^2)$ |
|---|---|---|---|---|---|---|
| Ni1 | 4h |  | 0.5 | 0.25 | 0 | -0.004(1) |
|  |  | $A_1$ | 0.117(2) | 0 | 0 |  |
|  |  | $B_1$ | 0 | 0 | 0 |  |
|  |  | $A_2$ | 0.034(4) | 0 | 0 |  |
|  |  | $B_2$ | 0 | 0.028(2) | 0 |  |
| Mn1 | 2a |  | 0 | 0 | 0 | -0.009(2) |
|  |  | $A_1$ | 0.114(3) | 0 | 0 |  |
|  |  | $B_1$ | 0 | 0 | 0 |  |
|  |  | $A_2$ | 0.034(8) | 0 | 0 |  |
|  |  | $B_2$ | 0 | 0 | 0 |  |
| In1 | 2d |  | 0 | 0.5 | 0 | 0.045(3) |
| Mn2 | 2d |  | 0 | 0.5 | 0 | 0.045(3) |
|  |  | $A_1$ | 0.051(4) | 0 | -0.015(6) |  |
|  |  | $B_1$ | 0 | 0 | 0 |  |
|  |  | $A_2$ | -0.021(5) | 0 | 0 |  |
|  |  | $B_2$ | 0 | 0 | 0 |  |

**(II)   Structural parameters obtained from the Rietveld refinement of SXRPD data considering distance constraint:**

**Table S2:** Atomic positions (x, y, z), amplitudes ($A_1$, $B_1$, $A_2$, $B_2$) of the modulation function, and atomic displacement parameter ($U_{iso}$) of the modulated martensite phase of $Ni_2Mn_{1.4}In0.6$ obtained from the Rietveld refinement using SXRPD data considering distance constraint.

| Atom | Wyckoff position | Modulation amplitude | x | y | z | $U_{iso}(Å^2)$ |
|---|---|---|---|---|---|---|
| Ni1 | 4h |  | 0.5 | 0.25 | 0 | 0.0123(19) |
|  |  | $A_1$ | 0.1001(3) | 0 | 0 |  |
|  |  | $B_1$ | 0 | 0 | 0 |  |
|  |  | $A_2$ | 0.0282(3) | 0 | 0 |  |
| | | B$_2$ | 0 | 0 | 0 | |
| Mn1 | 2a | | 0 | 0 | 0 | -0.0172(19) |
| | | A$_1$ | 0.1002(4) | 0 | 0 | |
| | | B$_1$ | 0 | 0 | 0 | |
| | | A$_2$ | 0.0285(5) | 0 | 0 | |
| | | B$_2$ | 0 | 0 | 0 | |
| In1 | 2d | | 0 | 0.5 | 0 | 0.030(3) |
| Mn2 | 2d | | 0 | 0.5 | 0 | 0.030(3) |
| | | A$_1$ | 0.1004(4) | 0 | 0.013(6) | |
| | | B$_1$ | 0 | 0 | 0 | |
| | | A$_2$ | 0.0286(5) | 0 | 0 | |
| | | B$_2$ | 0 | 0 | 0 | |

### (III) Rietveld refinements of neutron powder diffraction pattern at 300 K in the austenite phase

The results of Rietveld refinement using the neutron powder diffraction pattern of Ni$_2$Mn$_{1.4}$In$_{0.6}$ at 300 K (RT) in the austenite phase is shown in Fig.S1. The refinement was done by considering the atomic positions within the *Fm-3m* space group. The Ni and Mn atoms occupy the 8*c* (0.25 0.25 0.25) and 4*a* (0 0 0) Wyckoff positions, respectively, while In and extra Mn occupy the 4*b* (0.5 0.5 0.5) Wyckoff positions according to their relative occupancies.

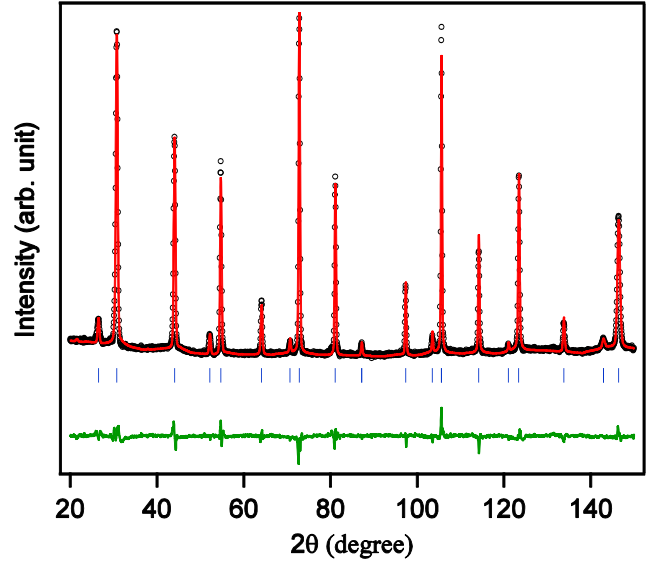

**Fig.S1**: (color online) Rietveld refinement of the neutron powder diffraction pattern at 300 K (austenite phase).

In the refinement, we also considered the possibility of anti-site disorder between different atoms (atomic sites) (e.g Ni(8*c*)-Mn(4*a*), Mn(4*a*)-In(4*b*) and Ni(8*c*)-In(4*b*)) but could not observe any improvement in the fits or the agreement (R) factors. Therefore the analysis of the Rietveld analysis of RT neutron diffraction data confirms absence of any substantial anti-site disorder in $Ni_2Mn_{1.4}In_{0.6}$ alloy.

**(IV) Rietveld refinements of neutron powder diffraction pattern in the martensite phase**

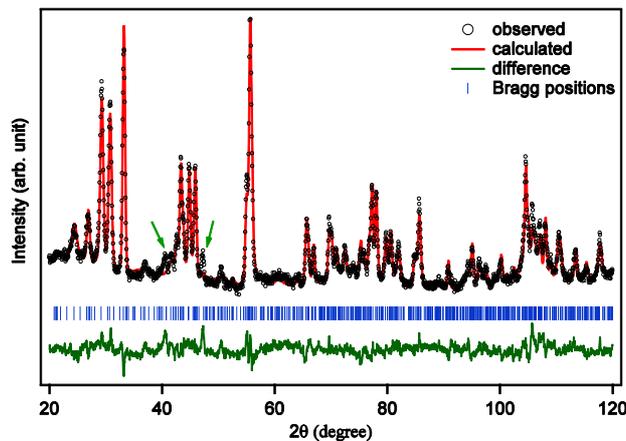

**Fig.S2**: (color online) Observed and calculated neutron diffraction pattern in the martensite phase (3K) using non-uniform displacement model. Green arrows indicates peaks due to Cryo-furnace wall material (Al).

After confirming the absence of any descernible anti-site disorder from the analysis of the RT neutron powder diffraction, we proceed to discuss the refinement of the structure of the martensite phase. The neutron diffraction data for the martensite phase was collected at the lowest possible temperature (3K). To investigate the modulated structure, we employed superspace (3+1) D formalism as for the SXRPD data. In the first step of Rietveld refinement, the refinement was carried out without any constraints on the amplitude or direction of atomic displacements for the atomic modulation functions of the different atoms, similar to that used for the analysis of the SXRPD pattern. The observed and calculated peak profiles are shown in Fig. S2 and corresponding refined structural parameters are given in Table S3.

It is interesting to note that while the refinement converged for a non-uniform atomic displacement model, but it led to unreasonable interatomic distances, as shown in Fig.S4a for some selected atomic pairs obtained using powder neutron diffraction. The interatomic distances given in Fig.S4a obtained from a non-uniform displacement model clearly indicates that the use of powder neutron diffraction also cannot resolve the issue of implausible interatomic distances and the problem lies with the modulation model itself, in agreement with SXRPD analysis. In the next step, we therefore considered a uniform displacement model for Rietveld refinement (Fig.S3) in which the amplitude of modulation for all the atomic sites was constrained to be identical (Table S4). The derived interatomic distances from the uniform atomic displacement model shown in the Fig.S4b clearly reveals that this model gives physically realistic interatomic distances that are acceptable for the shape memory Heusler compounds/alloys. Thus, our results based on both PND and SXRPD data reveal that the modulation in the martensite phase of $Ni_2Mn_{1.4}In_{0.6}$ involves uniform displacement of atoms and is, therefore, consistent with the predictions of the adaptive modulation model.

**Table S3:** Atomic positions (x, y, z), amplitudes ($A_1$, $B_1$, $A_2$, $B_2$) of the modulation function, and atomic displacement parameter ($U_{iso}$) of the modulated martensite phase of $Ni_2Mn_{1.4}In_{0.6}$ obtained from Rietveld refinement using neutron powder diffraction data for the non-uniform displacement model.

| Atom | Wyckoff position | Modulation amplitude | x | y | z | $U_{iso}(Å^2)$ |
|---|---|---|---|---|---|---|
| Ni1 | 4h |  | 0.5 | 0.25 | 0 | -0.0070(6) |
|  |  | $A_1$ | 0.1278(7) | 0 | 0 |  |
|  |  | $B_1$ | 0 | 0.011(2) | 0 |  |
|  |  | $A_2$ | 0.0223(13) | 0 | -0.011(1) |  |
|  |  | $B_2$ | 0 | -0.022(1) | 0 |  |
| Mn1 | 2a |  | 0 | 0 | 0 | -0.0070(6) |
|  |  | $A_1$ | 0.103(6) | 0 | -0.030(5) |  |
|  |  | $B_1$ | 0 | 0 | 0 |  |
|  |  | $A_2$ | 0.095(5) | 0 | 0 |  |
|  |  | $B_2$ | 0 | 0 | 0 |  |
| In1 | 2d |  | 0 | 0.5 | 0 | -0.0070(6) |
| Mn2 | 2d |  | 0 | 0.5 | 0 | -0.0070(6) |
|  |  | $A_1$ | 0.14(2) | 0 | -0.07(3) |  |
|  |  | $B_1$ | 0 | 0 | 0 |  |
|  |  | $A_2$ | 0.15(2) | 0 | -0.07(2) |  |
|  |  | $B_2$ | 0 | 0 | 0 |  |

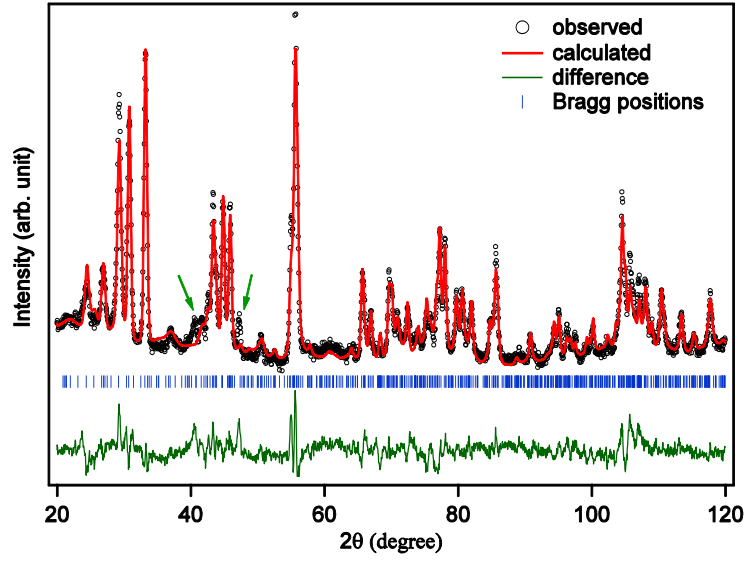

**Fig.S3**: (color online) Rietveld refinement of neutron powder diffraction pattern in the martensite phase (3K) considering the adaptive modulation model. Green arrows indicates peaks due to Cryo-furnace wall material (Al).

**Table S4 :** Atomic positions ($x, y, z$), atomic displacement parameter ($U_{iso}$) and amplitudes ($A_1$, $B_1$, $A_2$, $B_2$) of the modulation function of the modulated martensite phase of $Ni_2Mn_{1.4}In0.6$ obtained from Rietveld refinement using neutron powder diffraction data for the uniform displacement model.

| Atom | Wyckoff position | Modulation amplitude | x | y | z | $U_{iso}(Å^2)$ |
|---|---|---|---|---|---|---|
| Ni1 | 4h | | 0.5 | 0.25 | 0 | 0.0002(5) |
| Mn1 | 2a | | 0 | 0 | 0 | 0.0002(5) |
| In1 | 2d | | 0 | 0.5 | 0 | 0.0002(5) |
| Mn2 | 2d | | 0 | 0.5 | 0 | 0.0002(5) |
| | | $A_1$ | 0.1275(9) | 0 | 0.005(1) | |
| | | $B_1$ | 0 | 0 | 0 | |
| | | $A_2$ | 0.0377(17) | 0 | 0.004(2) | |
| | | $B_2$ | 0 | 0 | 0 | |

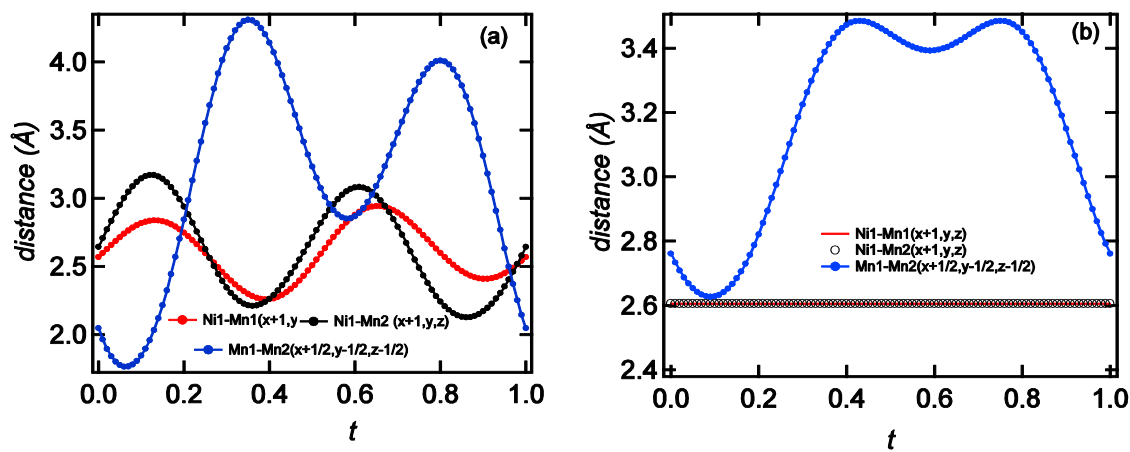

**Fig.S4**: (color online) Distance (selected) as a function of *t* parameters derived from (a) Non-uniform atomic displacement model (electronic instability model) showing unphysical values (less than 2 Å) and (b) Uniform atomic displacement model (adaptive phase model) showing values that are expected for these kind of intermetallic compounds/alloys.

**Rational approximant 3D superstructures obtained from the (3+1) D analysis of neutron diffraction data:**

It is possible to obtain the 3D superstructure from the analysis of diffraction data using (3+1) D superspace group. The refined structural parameters for the 3D rational approximant superstructures, which are obtained from the (3+1) D superspace group analysis of neutron powder diffraction data at 3K are shown in Table S5 for non-uniform displacement (electronic instability) model and in Table S6 for uniform displacement (adaptive phase) model.

**Table S5:** Lattice parameters, space group and atomic positions of 3D rational approximate structure of $Ni_2Mn_{1.4}In_{0.6}$ derived from (3+1) D incommensurate structure for non-uniform displacement (electronic instability) model.

| Crystal system: Monoclinic Space Group: P2/m Lattice parameters: a=4.3983(2) Å, b=5.5858(2) Å, c=12.9823(2) Å, β=92.796(9)° | | | | | |
|---|---|---|---|---|---|
| Atoms | Wyckoff position | x | y | z | Occ. |
| Mn1 | 1a | 0 | 0 | 0 | 1 |
| Mn2 | 1h | 1/2 | 1/2 | 1/2 | 1 |
| Mn3 | 2m | 0.00704 | 0 | 0.32342 | 1 |
| Mn4 | 2n | 0.67195 | 1/2 | 0.15906 | 1 |
| In1 | 1b | 0 | 1/2 | 0 | 0.6 |
| In2 | 1g | 1/2 | 0 | 1/2 | 0.6 |
| In3 | 2m | 0.75251 | 0 | 0.12560 | 0.6 |
| In4 | 2n | 0.99431 | 1/2 | 0.33461 | 0.6 |
| Mn5 | 1b | 0 | 1/2 | 0 | 0.4 |
| Mn6 | 1g | 1/2 | 0 | 1/2 | 0.4 |
| Mn7 | 2m | 0.75251 | 0 | 0.12560 | 0.4 |
| Mn8 | 2n | 0.99431 | 1/2 | 0.33461 | 0.4 |
| Ni1 | 2j | 1/2 | 1/4 | 0 | 1 |
| Ni2 | 2k | 0 | 3/4 | 1/2 | 1 |
| Ni3 | 4o | 0.59144 | 1/4 | 0.33676 | 1 |
| Ni4 | 4o | 0.13000 | 3/4 | 0.16365 | 1 |

**Table S6:** Lattice parameters, space group and atomic positions of 3D rational approximate structure of $Ni_2Mn_{1.4}In_{0.6}$ derived from (3+1) D incommensurate structure for uniform displacement (adaptive phase) model.

| Crystal system: Monoclinic Space Group: P2/m Lattice parameters: a=4.3983(2) Å b=5.5858(2) Å c=12.9823(2) Å β=92.796(9)° | | | | | |
|---|---|---|---|---|---|
| Atoms | Wyckoff position | x | y | z | Occ. |
| Mn1 | 1a | 0 | 0 | 0 | 1 |
| Mn2 | 1h | 1/2 | 1/2 | 1/2 | 1 |
| Mn3 | 2m | 0.07781 | 0 | 1/3 | 1 |
| Mn4 | 2n | 0.64303 | 1/2 | 0.16954 | 1 |
| In1 | 1b | 0 | 1/2 | 0 | 0.6 |
| In2 | 1g | 1/2 | 0 | 1/2 | 0.6 |
| In3 | 2m | 0.64303 | 0 | 0.16954 | 0.6 |
| In4 | 2n | 0.07781 | 1/2 | 1/3 | 0.6 |
| Mn5 | 1b | 0 | 1/2 | 0 | 0.4 |
| Mn6 | 1g | 1/2 | 0 | 1/2 | 0.4 |
| Mn7 | 2m | 0.64303 | 0 | 0.16954 | 0.4 |
| Mn8 | 2n | 0.07781 | 1/2 | 1/3 | 0.4 |
| Ni1 | 2j | 1/2 | 1/4 | 0 | 1 |
| Ni2 | 2k | 0 | 3/4 | 1/2 | 1 |
| Ni3 | 4o | 0.57781 | 0.25000 | 0.33376 | 1 |
| Ni4 | 4o | 0.14303 | 0.75000 | 0.16954 | 1 |

**(V)    High resolution Transmission electron microscopy  (HRTEM):**

To further support our results obtained from the diffraction studies about the uniform displacement model as origin of modulation in $Ni_2Mn_{1.4}In_{0.6}$, we performed HRTEM studies. Fig. S5 shows the electron diffraction pattern at 300 K. The diffraction patterns clearly confirm the austenite structure of $Ni_2Mn_{1.4}In_{0.6}$ at 300 K, which is also in agreement with the magnetization data (shown in the main manuscript).  The fast Fourier transform (FFT) of HRTEM image shown in Fig.5a of the main text   for [210] zone in the martensite phase is shown in Fig.S6a  while the simulated diffraction pattern of [210] zone  using the positional coordinates for the 3D rational approximant superstructure (Table S6) of adaptive phase modulation model is shown in Fig.S6b. Both show excellent match and thus the HRTEM images also confirm the uniform displacement model.

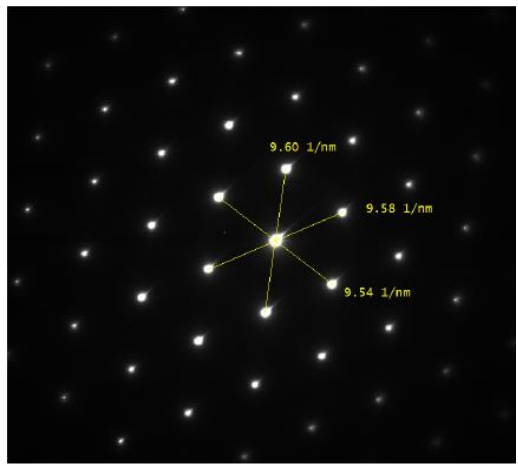

**Fig. S5:**  Diffraction pattern in the austenite phase (room temperature) of $Ni_2Mn_{1.4}In_{0.6}$ for <111>  zone confirms the cubic structure.

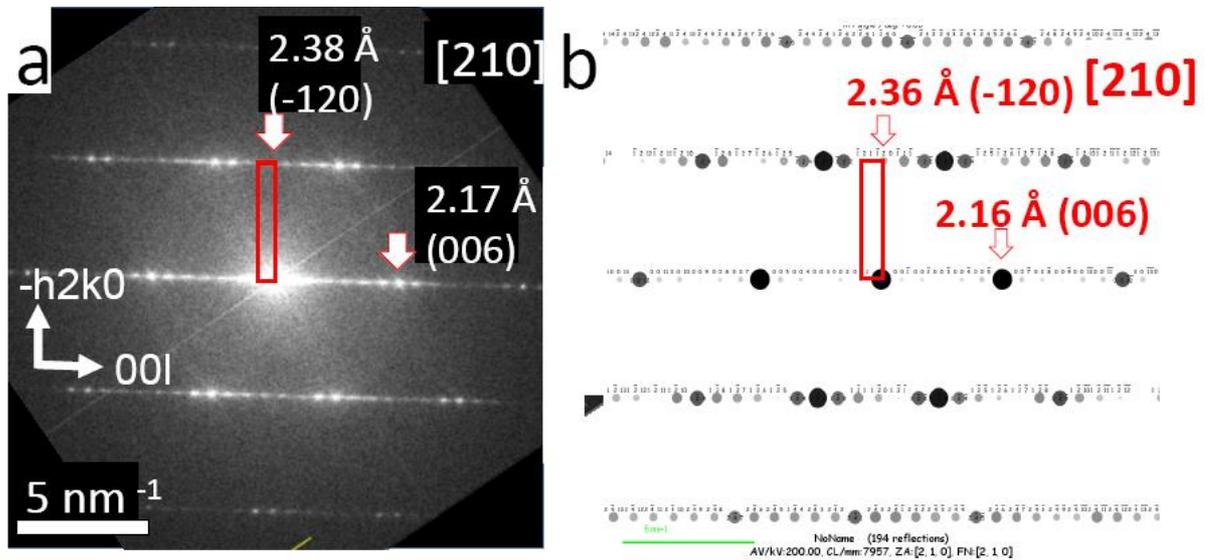

**Fig.S6.** (a) FFT of high-resolution image of $Ni_2Mn_{1.4}In_{0.6}$, recorded from [210] zone. (b) Simulated diffraction pattern of [210] zone showing expected intensity distribution of reflections.

## B. Theoretical calculations:

### a. Band structure and ground state magnetism:

In order to verify the results of Rietveld refinements and HRTEM studies, we also performed first-principles calculations for martensite phase for both adaptive phase model and non-uniform displacement modulation model using the fully relativistic Green function SPR-KKR package [6] and employing the CPA alloy theory to account for the chemical Mn-In disorder. The exchange-correlation potential was treated within the conventional local density approximation [7]. For first-principles calculations 3D rational approximant superstructures for both non-uniform displacement (Table S5) and adaptive phase (Table S6) models were used. For non-uniform displacement model, many quantities (such as the Fermi energy, local magnetic moments, etc.) appeared to be unrealistic. In particular, the total energy appears to be incomparably high (several hundred Ry per unit cell) with respect to that of the adaptive phase model. Thus, the non-uniform displacement model appears to be unrealistic and further we consider only the AM-phase.

By going from the high symmetric Bain distorted tetragonal martensite (SG 139) to AM-phase (SG 10), its high-symmetry Wyckoff positions split into lower-symmetric sites (e.g. 2*a* Wyckoff site in SG 139 occupied by Mn splits into 1*a*, 1*h*, 2*m* and 2*n* in SG 10), but their environment still remains to be rather similar, by resulting in a very similar DOS (compare, for instance, partial DOS of Mn in 1*a* and 1*h* (SG 10) with that of 2*a* (SG 139) in Fig. S7a and b). On the other hand, these slight changes in the partial DOS from inequivalent sites lead to a more complicated band structure of the adaptive phase model compared to the tetragonal martensite. It also exhibits more disorder broadening since the random (random site occupation by Mn and In) positions 1*b*, 1*g*, 2*m* and 2*n* produce the same broadened spectra as 2*b* in SG 139, but shifted in energy with respect to one another. In particular, this is also reflected in the form of majority-spin peak (~-0.75 eV below $E_F$) of Ni states (2*j* and 4*o*), which has a more pronounced single-Lorentzian form compared to that of 4*d* in SG 139. Furthermore, comparison of the total energies of the adaptive phase model with respect to the magnetization direction reveals an easy magnetic anisotropy axis along the *b*-side: $E(c) - E(b) \approx$ 0.9 and $E(a) - E(b) \approx$ 1.2 meV/u.c., which is compatible with the results of the neutron refinement. Thus the first principle calculations also support the uniform displacement model.

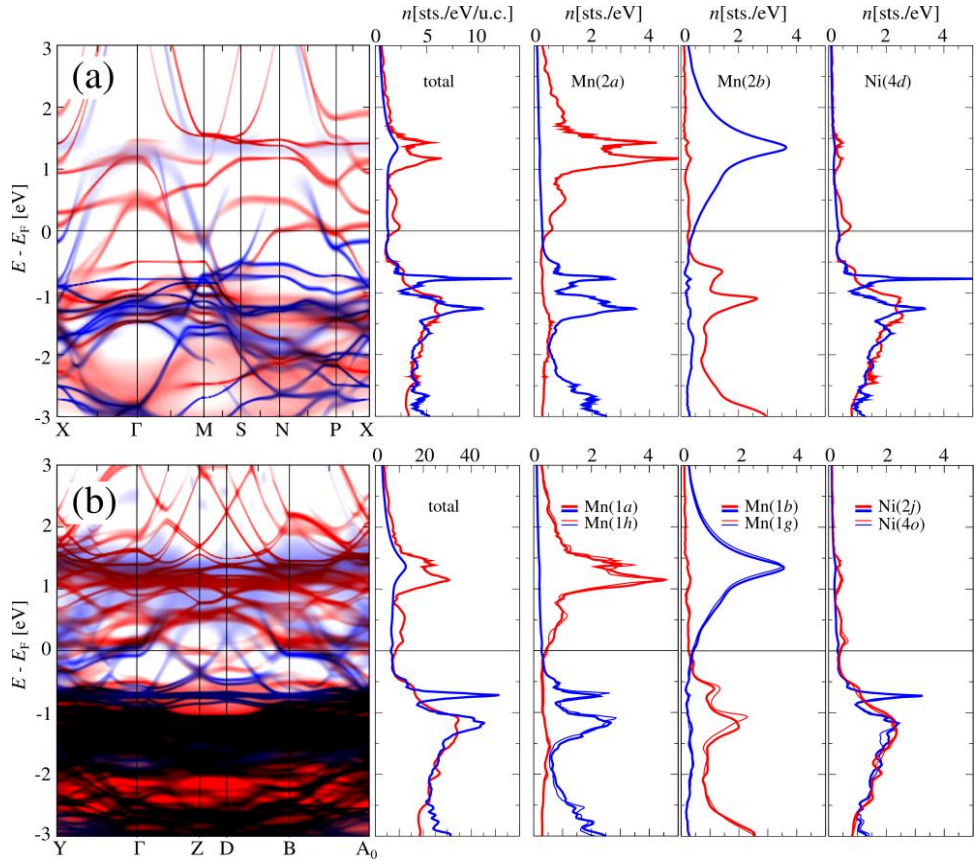

**Fig.S7**. Electronic structure, total and type-resolved densities of states of $Ni_2Mn_{1.4}In_{0.6}$ in the (a) tetragonal martensite phase (SG 139), (b) modulated martensite within the adaptive phase model (SG 10). Blue and red colors mark the majority- and minority-spin states. Since in case (b) there are more inequivalent positions compared to (a), the DOS for some of them is shown on the same plot (e.g. Mn(1$a$) with thick and Mn(1$h$) with thin lines).

*sanjay.singh@cpfs.mpg.de